\documentclass[10pt]{IEEEtran}
\nonstopmode
\usepackage{amsmath}
\usepackage{amsfonts}
\usepackage{fullpage}
\usepackage{graphicx}
\usepackage{url}
\usepackage{float}
\usepackage{lipsum}
\usepackage[margin=0.75in,footskip=0.25in]{geometry}

\newcommand\blfootnote[1]{%
	\begingroup
	\renewcommand\thefootnote{}\footnote{#1}%
	\addtocounter{footnote}{-1}%
	\endgroup
}

\newcommand\zfrac[2]{\leavevmode\kern.1em\raise.5ex\hbox{\the\scriptfont0
#1}\kern-.1em/\kern-.15em\lower.25ex\hbox{\the\scriptfont0 #2}}

\newcommand\zline{\\ \hline}
\newcommand\zleft[1]{\multicolumn{1}{|l|}{#1}}
\newcommand\zcentB[1]{\multicolumn{1}{|c|}{#1}}
\newcommand\zcentR[1]{\multicolumn{1}{c|}{#1}}

\newcommand\ZbeginE{ \begin{enumerate} }
\newcommand\ZendE{     \end{enumerate} }
\newcommand\ZbeginI{ \begin{itemize} }
\newcommand\ZendI{     \end{itemize} }

\begin{document}

\author{
\begin{tabular}[t]{c@{\extracolsep{8em}}c} 
Amanda Metzner                & Daniel Zwillinger     \\
BAE Systems FAST Labs         & BAE Systems FAST Labs \\ 
600 District Ave.             & 600 District Ave.     \\
Burlington, MA 01803          & Burlington, MA 01803  \\
amanda.metzner@baesystems.com & daniel.zwillinger@baesystems.com \\
\end{tabular}
}

\title{Kuhn Poker with Cheating and Its Detection}

\maketitle

\begin{abstract}
  Poker is a multiplayer game of imperfect information and has been
  widely studied in game theory.  Many popular variants of poker
  (e.g., Texas Hold'em and Omaha) at the edge of modern game theory
  research are large games. However, even toy poker games, such as
  Kuhn poker, can pose new challenges.  Many Kuhn poker variants have
  been investigated: varying the number of players, initial pot size,
  and number of betting rounds. In this paper we analyze a new variant
  -- Kuhn poker with cheating and cheating detection. We determine how
  cheating changes the players' strategies and derive new analytical
  results.
\end{abstract}


\blfootnote{Approved for public release; unlimited distribution.}
\blfootnote{Not export controlled per ES-FL-102220-0153.}

\section{Introduction}
\label{ssec:Introduction}
Poker is one of the most popular games studied among game theory
researchers. The complex nature of the game provides challenging
problems, however its complexity and size also make poker a difficult
game to study. Kuhn poker, a toy poker game, can pose theoretical
challenges at a much more manageable size.
Multiple variants of Kuhn poker have previously been studied and
optimal strategies determined; varying the number of players, the
number of betting rounds, the initial pot size, etc \cite{Billingham}.
A variant that has not been studied previously is the incorporation
of cheating and its detection.
What happens when one or both players cheat? How does cheating alter
the players' strategies?

\section{Preliminaries}
\label{ssec:Preliminaries}

\subsection{Game Theory}
Game theory is a theoretical framework for comprehending social
interactions among competing players~\cite{gametheory}.
In some regards, game theory is the science of strategy: the optimal
decision-making of independent and competing agents in a strategic
setting.
In game theory, the game itself serves as a model of an interactive
decision making among rational players.

Game theory terms:
\ZbeginI
\item Players: The game decision-maker
\item Strategy: A player's plan of action
\item Payoff: The player payouts received from a particular outcome
\item Information set: The information available to a player at a game decision point
\ZendI

In this paper we study Kuhn poker with the incorporation of cheating
and its detection.
The goal is to determine how cheating changes the players' strategies.

\subsection{Kuhn Poker}
The rules of Kuhn poker are simple.
There are two players using a three card deck (the cards are called K, Q, and J for King, Queen, and Jack).
Each player antes~\$1 at the beginning of the game.
Each player
is dealt one card.
Player 1 goes first.
There is a single betting
round, with a maximum of one bet of~\$1.
If no player folds, then the player with the higher card wins (K beats
Q and J, while Q beats J).

Kuhn poker is asymmetric, the first player has a disadvantage with an
expected payoff~\cite{kuhnpoker} (under optimal play by both sides) of
$-\zfrac{1}{18}$ dollars.
Section~\ref{ssec:AnalyticalComp} explains this in more detail.

Note that optimal play of classical Kuhn power incorporates bluffing.
For example, in Kuhn poker (with no cheating), if player~1 is dealt a
J then, with some probability, player~1 bets even though
they will lose to whatever card player~2 has.
The reason is simple, if player~1 bets with a J, and player 2 has a Q,
then player~2 may “think” player~1 has a K, and fold.

\section{Incorporating Cheating and Its Detection}
\label{ssec:Incorporating}
There are many ways in which cheating could occur in Kuhn poker.
For example, a player could shuffle the deck, or otherwise arrange it,
in such a way that all 6 two-card distributions are not equally likely.
The cheating variation that we choose to analyze was having either player, or
both players, look at the third/face-down card.

If either player looks at the face-down card, then that player knows
who has the best hand.
For the cheating player, it is a game of complete information.
Note that cheating alone will not cause a player to win a hand – they
may have a losing hand (e.g., the~J).
Note that bluffing against a cheating player will not be successful.

We analyzed several variations of our way of cheating:
\ZbeginE
  \item Classic (no cheating by either player)
  \item Player~1 (alone) cheats probabilistically
  \item Player~2 (alone) cheats probabilistically
  \item Players 1 and 2 both cheat probabilistically
  \item Players 1 and 2 both cheat probabilistically and they both
    detect cheating probabilistically.
\ZendE

Probabilistic detection of cheating was also incorporated into the
game. For example, ``Only 10\% of the time will player~1 determine if
player~2 has cheated."
The reason for probabilistic detection is that there could be a cost
associated with detection, so a player may only check for cheating some of
the time.
For a game that incorporates cheating detection, the rules
differ slightly:
\ZbeginI
  \item If \textit{one player cheats and the other player detects it},
    then the round is played out per usual with betting and folding
    and then the \textit{detector wins}, independent of the cards
    dealt or how the game played out.
  \item If \textit{both players cheats and both players detect that
    the other is cheating}, then the round is played out per usual
    with betting and folding and then \textit{neither player wins},
    independent of the cards dealt or how the game played out. In this
    case the monies bet are returned.

    \ZendI

\section{Game Computation}
\subsection{Kuhn Poker Analytical Computation}
\label{ssec:AnalyticalComp}
For classical Kuhn poker there are a range of optimal strategies for
player~1, all yielding the same optimal result~\cite{kuhnpoker}.
While the strategy for player~2 is fixed, the optimal strategy for
player~1 depends on a parameter "$a$", which can have any value
between 0 and $\zfrac{1}{3}$, inclusive.

If player~1 is dealt a J, then player~1 bets (bluffs) with probability
$a$; if player~1 is dealt a K, then player~1 bets with
probability~$3a$.
A complete description of both players’ optimal strategy is contained
in Tables~\ref{table:stratP1} and~\ref{table:stratP2}; we call this
the ``fair strategy.''

\begin{table}[t!]
\caption{Optimal play strategies for player~1 when both players play
  fairly (no cheating)}
\centering
\begin{tabular}{c|c|c|c|c|}
\cline{2-5} & \multicolumn{4}{c|}{player~1 \dots} \\
\cline{2-5} & \multicolumn{2}{c|}{\dots initial
  play} & \multicolumn{2}{c|}{\dots response to a bet}\zline
\zleft{card dealt} & \zleft{prob bet}  & \zleft{prob check}
                   & \zleft{prob call} & \zleft{prob fold} \zline
\zcentB{K} & $3a$ & $1-3a$ & 1       & 0       \zline
\zcentB{Q} & 0    & 1      & $1/3+a$ & $2/3-a$ \zline
\zcentB{J} & $a$  & $1-a$  & 0       & 1       \zline
\end{tabular}
\label{table:stratP1}
\end{table}

\begin{table}[t!]
\caption{Optimal play strategies for player~2 when both players play
  fairly (no cheating)}
\centering
\begin{tabular}{c|c|c|c|c|}
\cline{2-5} & \multicolumn{4}{c|}{player~2 response to a \dots} \\
\cline{2-5} & \multicolumn{2}{c|}{\dots check} & \multicolumn{2}{c|}{\dots bet}\zline
\zleft{card dealt} & \zleft{prob bet} & \zleft{prob call}
                  & \zleft{prob call} & \zleft{prob fold} \zline
\zcentB{K} & 1 & 0 & 1 & 0 \zline
\zcentB{Q} & 0 & 1 & $\zfrac{1}{3}$ & $\zfrac{2}{3}$ \zline
\zcentB{J} & $\zfrac{1}{3}$ & $\zfrac{2}{3}$ & 0 & 1 \zline
\end{tabular}
\label{table:stratP2}
\end{table}

Table~\ref{table:fairplay} shows that the payoff to player~1 using the optimal
strategy (including the value "$a$") is $-\zfrac{1}{18}$.
Leftmost in Table~\ref{table:fairplay} are the 6 equally possible ways
that two cards can be dealt.
Adjacent to that are the probabilistic amounts that each player can
expect to win for that deal.
We presume the first player never folds as their initial play, there
is no benefit to this.
The result shows that player~1 expects to win $-\zfrac{1}{18}$
(that is, a loss) each play of the game, and the result does not
depend on the value of~``$a$.''

\begin{table}[t!]
  \caption{The value of the game to player~1 when both players are playing fairly}
  \centering
\begin{tabular}{|c|ccc}
\cline{1-2} \multicolumn{2}{|c|}{{cards dealt}} & \multicolumn{1}{l}{}
& \multicolumn{1}{l}{}\zline {player~1} &
\zcentR{{player~2}} & \zcentR{{player~1 winnings}} & \zcentR{{player~2 winnings}}\zline
K & \zcentR{J} & \zcentR{$\zfrac{2}{9} -\zfrac{a}{6}$} & \zcentR{}\zline
K & \zcentR{Q} & \zcentR{$\zfrac{1}{6} +\zfrac{a}{6}$} & \zcentR{}\zline
Q & \zcentR{J} & \zcentR{$\zfrac{4}{27}+\zfrac{a}{9}$}
               & \zcentR{$\zfrac{1}{27}-\zfrac{a}{18}$}\zline
Q & \zcentR{K} & \zcentR{} & \zcentR{$\zfrac{2}{9}+\zfrac{a}{6}$}\zline
J & \zcentR{K} & \zcentR{} & \zcentR{$\zfrac{1}{6}+\zfrac{a}{6}$}\zline
J & \zcentR{Q} & \zcentR{$\zfrac{a}{9}$} & \zcentR{$\zfrac{1}{6}-\zfrac{a}{18}$}\zline
\multicolumn{4}{|l|}{net winnings to player~1 \quad $-\zfrac{1}{18}$}\zline
\end{tabular}
\label{table:fairplay}
\end{table}

Games 2 and 3 (defined in the last section) have one player cheating
and the other player unaware of this cheating.
In this case the natural question is ``How much does the cheating player benefit?''
The analysis is shown in
\ZbeginI
\item Table~\ref{table:P2C}, where player~2 cheats 100\% of the time, and player~1 plays their fair strategy
\item Table~\ref{table:P1C}, where player~1 cheats 100\% of the time, and player~2 plays their fair strategy
\ZendI
where new optimal strategies were determined for the cheating player.

\begin{table}[t!]
\caption{The value of the game to player~1 when player~2 cheats and
  player~1 plays fairly}
\centering
\begin{tabular}{|c|ccc}
\cline{1-2} \multicolumn{2}{|c|}{{cards dealt}} & \multicolumn{1}{l}{}
& \multicolumn{1}{l}{}\zline
      {player~1} &
\zcentR{{player~2}} & \zcentR{{player~1
    winnings}} & \zcentR{{player~2 winnings}}\zline
K & \zcentR{J} & \zcentR{$\zfrac{1}{6}$} & \zcentR{}\zline
K & \zcentR{Q} & \zcentR{$\zfrac{1}{6}$} & \zcentR{}\zline
Q & \zcentR{J} & \zcentR{$\zfrac{1}{9}+\zfrac{a}{9}$} & \zcentR{$\zfrac{1}{9}-\zfrac{a}{6}$}\zline
Q & \zcentR{K} & \zcentR{} & \zcentR{$\zfrac{2}{9}+\zfrac{a}{6}$}\zline
J & \zcentR{K} & \zcentR{} & \zcentR{$\zfrac{1}{6}$}\zline
J & \zcentR{Q} & \zcentR{} & \zcentR{$\zfrac{1}{6}$}\zline
\multicolumn{4}{|l|}{net winnings to player~1 \quad $-\zfrac{2}{3}-\zfrac{a}{9}$}\zline
\end{tabular}
\label{table:P2C}
\end{table}

\begin{table}[t!]
\caption{The value of the game to player~1 when player~1 cheats and
  player~2 plays fairly}
\centering
\begin{tabular}{|c|ccc}
\cline{1-2} \multicolumn{2}{|c|}{{cards dealt}} & \multicolumn{1}{l}{}
& \multicolumn{1}{l}{}\zline
{player~1} & \zcentR{{player~2}} & \zcentR{{player~1 winnings}} & \zcentR{{player~2 winnings}}\zline
K & \zcentR{J} & \zcentR{$\zfrac{2}{9}$} & \zcentR{}\zline
K & \zcentR{Q} & \zcentR{$\zfrac{2}{9}$} & \zcentR{}\zline
Q & \zcentR{J} & \zcentR{$\zfrac{2}{9}$} & \zcentR{}\zline
Q & \zcentR{K} & \zcentR{} & \zcentR{$\zfrac{1}{6}$}\zline
J & \zcentR{K} & \zcentR{} & \zcentR{$\zfrac{1}{6}$}\zline
J & \zcentR{Q} & \zcentR{$\zfrac{1}{9}$} & \zcentR{$\zfrac{1}{18}$}\zline
\multicolumn{4}{|l|}{net winnings to player~1 \quad $\zfrac{7}{18}$}\zline
\end{tabular}
\label{table:P1C}
\end{table}

When player~2 cheats then player~1 expects to win
$-\zfrac{2}{3}-\zfrac{a}{9}$ (that is, a loss, for any value of "$a$")
on each play of the game.
While the value of "$a$" did not make a difference in the fair game
(with neither side cheating), it does make a difference when player~2
cheats.
Hence, to minimize the loss against a possibly cheating opponent,
player~1 should always play their optimal strategy with $a=0$.
When using $a=0$, player~1 wins $-\zfrac{2}{3}$ (that is, a loss).
This is a change (loss) of~$\zfrac{11}{18}$ compared to the fair game.

Alternately, player~1 could cheat and player~2 could be unaware of
this.  In this case (Table~\ref{table:P1C}), player~1 now expects to
win $\zfrac{7}{18}$ (that is, a gain) each play of the game.
This is a change (increase) of~$\zfrac{8}{18}$ compared to the fair game.

We conclude that player~2 is more motivated to cheat than player~1 is.
That is, the benefit to player~2, when player~2 cheats, exceeds the
benefit to player~1, when player~1 cheats.

\subsection{Kuhn Poker Programmatic Computation}
\label{ssec:ProgramComp}
\begin{figure}[ht!]
  \centering
  \includegraphics[width=3.5in]{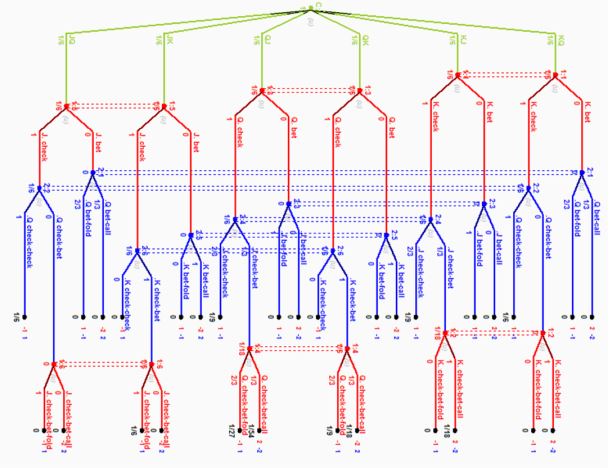}
  \caption{Game tree representation in Gambit}
  \label{fig:gamegraph}
\end{figure}

The optimal strategy for Kuhn poker can also be solved
programmatically using software tools such as Gambit~\cite{gambit}.
Gambit is an open source library of game theory tools for the
construction and analysis of finite extensive and strategic games.
The Gambit GUI displays the extensive tree form of the game as shown
in Figure~\ref{fig:gamegraph}.
Some information about this tree:
\ZbeginI
  \item The player actions are shown on the tree edges.
  \item The top/root node in green is a chance node, it has 6 links
    representing the 6 possible card dealings.
  \item Player~1 actions are in red, player~2’s are in blue.
  \item Figure~\ref{fig:gamegraph} displays the information sets (ISs) as dotted
    lines – these are the states that a player cannot distinguish
    between with the available knowledge.
  \item Kuhn poker has 24 decision nodes, of which 12 are distinct
after IS consideration.
\ZendI

After the game was created in extensive form, Gambit solved it.
In the fair case (no cheating), the result are the same as the solutions previously shown in
Table~\ref{table:stratP1}, with "$a$" equal to zero.

\begin{figure}[ht!]
  \centering
  \includegraphics[width=2.5in]{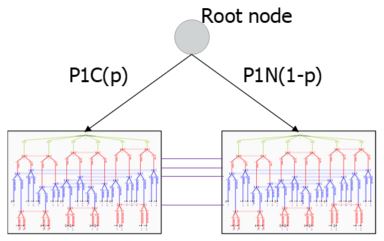}
  \caption{Representation of Kuhn poker when player~1 cheats
    probabilistically}
  \label{fig:KP_1PC}
\end{figure}

The way to implement our extensions to the classic version of Kuhn
poker is to build on the tree in Figure~\ref{fig:gamegraph}.
Specifically, a game in which only player~1 cheats probabilistically
is shown in Figure~\ref{fig:KP_1PC}. For this Figure:
\ZbeginI
  \item P1C is the path from the root node when player~1 does cheats
    (with probability $p$).
  \item P1N is the path from the root node when player~1 does not
    cheats (with probability $1-p$).
  \item Each boxed images at the bottom of the tree in
    Figure~\ref{fig:KP_1PC} represents all of
    Figure~\ref{fig:gamegraph}.

\ZendI

There are lines drawn between the boxed images, these represent
the IS connections.
For example, player~2, who is not cheating, cannot determine from the
information available to them if they are on the left or right branch
of the tree in Figure~\ref{fig:KP_1PC}.
Hence, the actions taken by player~2 when player~1 bets, cannot be
different.
Stated differently, player~2 does not know if player~1 is cheating or
not.

\begin{figure}[ht!]
  \centering
  \includegraphics[width=3.5in]{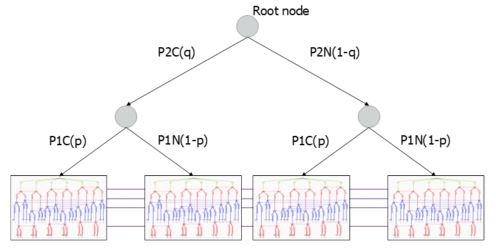}
  \caption{Representation of Kuhn poker when players 1 and 2 both
    cheat probabilistically}
  \label{fig:KP_2PC}
\end{figure}

The analogous game when only player~2 cheats is to replace the labels
in Figure~\ref{fig:KP_1PC} from “P1C($p$)” to “P2C($q$)” (that is,
player~2 cheats with probability~$q$) and from “P1N($1-p$)” to
“P2N($1-q$)”.
To implement a game where both players 1 and 2 cheat
probabilistically, we merely need to extend the height of the game
tree one more level, as shown in Figure~\ref{fig:KP_2PC}.

In this case, the ISs are connected across all four sub-figures of the
bottom layer, depending on what a player knows when it is their turn.
Note that the same result is obtained if the player~2 options are
followed by the player~1 options (as shown in Figure~\ref{fig:KP_2PC})
or if the order is reversed and the first options are for player~1.
Adding cheating detection is performed in the same
manner as adding cheating.

\begin{figure}[ht!]
  \centering
  \includegraphics[width=3.5in]{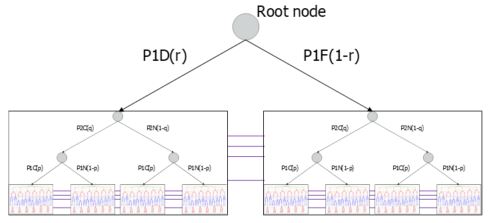}
  \caption{Representation of Kuhn poker when both players cheat and
    player~1 can detect cheating}
  \label{fig:KP_1PD}
\end{figure}

Figure~\ref{fig:KP_1PD} adds cheating detection by player~1.
In this case, player~1 detects cheating on the left branch (case “P1D”
which occurs with probability~$r$) and does not on the right branch
(case “P1F” which occurs with probability~$1-r$).
Once again, there are horizontal lines connecting the ISs.
Additionally, when we incorporated cheating detection, we also changed
the game payoffs.
We implemented the rule that a player caught cheating always lost,
even if the other player had the lower card.
When both players cheat, it is a draw.

\begin{figure}[ht!]
  \centering
  \includegraphics[width=3.5in]{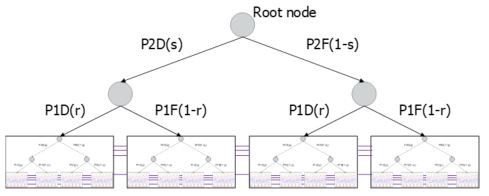}
  \caption{Representation of Kuhn poker when both players cheat and
    both players can detect cheating}
  \label{fig:KP_2PD}
\end{figure}

Unsurprisingly, we can also create the tree representing the game when
each player cheats probabilistically and each player can also
probabilistically detect cheating by the other player, as shown in
Figure~\ref{fig:KP_2PD}.
We solved this game using Gambit, there are approximately 1000 nodes
when both cheating and cheating detection were incorporated.

We used python to create the tree in Figure~\ref{fig:KP_2PD}, and
specify all the information sets.
Gambit determined the optimal game strategy and the corresponding
payoffs, using the linear program algorithm.

The Gambit calculations for the payoff to player~1, when only one
player cheats, are shown in Tables~\ref{table:P2CP1A}
and~\ref{table:P1CP2A}.
Unlike the analytical calculations shown in
Section~\ref{ssec:AnalyticalComp}, the programatic calculations
include players \textit{adapting} to cheating.
For example, player~1 may not know that player~2 is cheating, but
based on player~2's actions, player~1 changes their strategy with
respect to the fair play strategy.
Therefore, the incorporation of cheating also influences the
non-cheating player's game play.

\begin{table}[t!]
\caption{The value of the game to player~1 when player~2 cheats and
  player~1 adapts}
\centering
\begin{tabular}{|cccc}
\cline{1-2} \multicolumn{2}{|c|}{{cards dealt}} & \multicolumn{1}{l}{}
& \multicolumn{1}{l}{}\zline
\zcentB{{player~1}} &
\zcentR{{player~2}} & \zcentR{{player~1 winnings}} & \zcentR{{player~2 winnings}}\zline
\zcentB{K} & \zcentR{J} & \zcentR{1}  & \zcentR{}\zline
\zcentB{K} & \zcentR{Q} & \zcentR{1}  & \zcentR{}\zline
\zcentB{Q} & \zcentR{J} & \zcentR{$\zfrac{10}{9}$} & \zcentR{$\zfrac{1}{9}$}\zline
\zcentB{Q} & \zcentR{K} & \zcentR{}   & \zcentR{$\zfrac{5}{3}$}\zline
\zcentB{J} & \zcentR{K} & \zcentR{}   & \zcentR{1}\zline
\zcentB{J} & \zcentR{Q} & \zcentR{}   & \zcentR{1}\zline
\textbf{} & \textbf{} & \textbf{}   & \zcentR{\textbf{}}\zline
\multicolumn{2}{|l|}{{net winnings to player~1}} & \multicolumn{2}{c|}{{$-\zfrac{1}{9}$}}\zline
\end{tabular}
\label{table:P2CP1A}
\end{table}

\begin{table}[t!]
\caption{The value of the game to player~1 when player~1 cheats and
  player~2 adapts}
\centering
\begin{tabular}{|cccc}
\cline{1-2} \multicolumn{2}{|c|}{{cards dealt}} & \multicolumn{1}{l}{}
& \multicolumn{1}{l}{}\zline
  \zcentB{{player~1}}
& \zcentR{{player~2}}
& \zcentR{{player~1 winnings}}
& \zcentR{{player~2 winnings}}\zline
\zcentB{K} & \zcentR{J} & \zcentR{1}  & \zcentR{}\zline
\zcentB{K} & \zcentR{Q} & \zcentR{$\zfrac{5}{3}$} & \zcentR{}\zline
\zcentB{Q} & \zcentR{J} & \zcentR{1}  & \zcentR{}\zline
\zcentB{Q} & \zcentR{K} & \zcentR{}   & \zcentR{1}\zline
\zcentB{J} & \zcentR{K} & \zcentR{}   & \zcentR{1}\zline
\zcentB{J} & \zcentR{Q} & \zcentR{$\zfrac{1}{9}$} & \zcentR{$\zfrac{10}{9}$}\zline
\textbf{} & \textbf{} & \textbf{} & \zcentR{\textbf{}}\zline
\multicolumn{2}{|l|}{{net winnings to player~1}} & \multicolumn{2}{c|}{{$\zfrac{1}{9}$}}\zline
\end{tabular}
\label{table:P1CP2A}
\end{table}

When player~2 cheats and player~1 adapts to the cheating, player~1
expects to win $-\zfrac{1}{9}$, this is a change (loss)
of~$\zfrac{1}{18}$ compared to the fair game.
When player~1 cheats and player~2 adapts to the cheating, player~1
expects to wins $\zfrac{1}{9}$, this is a change (gain)
of~$\zfrac{3}{18}$ compared to the fair game.
In this case, when players adapt to cheating, player~1 is more
motivated than player~2 to cheat.
This finding is opposite of what was seen in
Section~\ref{ssec:AnalyticalComp}, where player~2 was more motivated
to cheat.

\section{Results}
We investigated the effects of cheating and its detection on the 5
games described in section~\ref{ssec:Incorporating}.
Game 1 is the classic, fair play, version of Kuhn poker.
This version has been well studied, and our numerical results
reproduce the known results for the parameter value  $a=0$.

Games 2--4 focus primarily on incorporating cheating into Kuhn poker.
We analyzed these cases for cheating probabilities ranging from 0
(never cheat) to 1 (always cheat) for each player.

\begin{figure}[t]
  \includegraphics[width=3in]{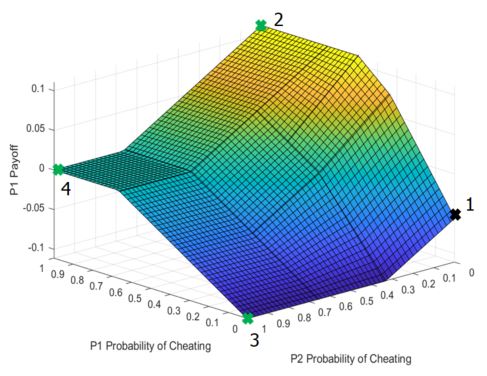}
  \caption{The value of the game to player~1 when both players 1 and 2 are
    probabilistically cheating}
  \label{fig:resultscheat}
\end{figure}

In this analysis, the game was fully adaptive; equivalently each player
knew the probability of the other player's cheating.
The computational results are shown in Figure~\ref{fig:resultscheat},
for which the axes are the probability that each player cheats and the
height is the expected payoff to player~1.

There are several observations:
\ZbeginE
  \item The payoff surface has several bi-linear patches – we have not
    yet been able to analytically explain this.
  \item The 4 marked corners are special cases, the values below are the expected payoff to player~1:
\ZbeginE
\item Point 1: fair game, neither player cheats: $–\zfrac{1}{18}$
\item Point 2: player~1 cheats, player~2 does not: $\zfrac{1}{9}$
\item Point 3: player~2 cheats, player~1 does not: $–\zfrac{1}{9}$
\item Point 4: both players cheat: $0$
\ZendE
  \item When both players cheat (Point 4), the payoff surface is flat
    and equal to zero for a range of cheat values for both players 1
    and~2.  For example, when both players cheat 90\% of the time, the
    expected payoff is zero.  It is surprising that if either player
    changes their likelihood of cheating from 90\% to 89\% or 91\%,
    there is no change in the payoff.  \ZendE

\begin{figure}[t]
  \includegraphics[width=3in]{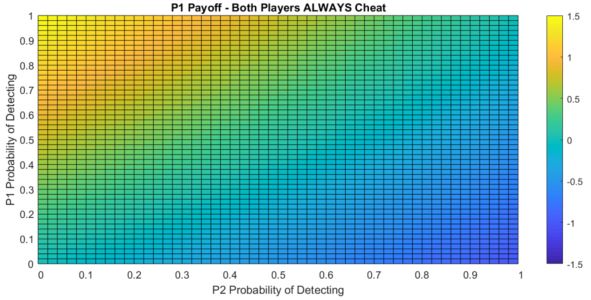}
  \caption{The value of the game to player~1 when both players are probabilistically
    detecting}
  \label{fig:detectCheatP1}
\end{figure}

Note that the payoff to player~1, when only one player cheats, is
different from the payoff in Section~\ref{ssec:AnalyticalComp}.
Section~\ref{ssec:AnalyticalComp} had one player cheating, while the
other naively played fairly.
In this case, there is full adaption to the cheating (``I know that
you are cheating $p$\% of the time.'')
Game 5 incorporated both cheating and detecting cheating.
We implemented multiple iterations of game 5,
varying the probability of cheating and the probability of detecting
cheating.
The version of most interest was when both players always cheated and
both players' detection probabilities varied from 0 to~1.
In this case, player 1 payoffs are shown in
Figures~\ref{fig:detectCheatP1}.
The axes represent the probability of each player detecting cheating
and the payoffs are color coded (higher payoffs = yellow, lower
payoffs = blue).
These figures show that player~1 benefits more from detecting cheating
than player~2 does. Even in a fair game, player~1 is at a disadvantage
by betting first. Therefore, player 1 has more to gain by detecting
cheating.

\section{Conclusion}

Kuhn poker is a toy poker game involving two players and three cards; it is zero sum.
Each player is dealt one card, while the third card is face down.
Classic Kuhn poker is a game of imperfect information, each player
only knowing their own card.
In this paper we analyzed Kuhn poker when the game allowed cheating;
one or both players peaked at the face down card.

In the analytical analysis, we assumed only one player was cheating
and the other player was using the ``fair'' (non-cheating) strategy;
this is the non-adaptive approach.
In this case player~2 was more motivated to cheat than player~1.
We also analyzed cheating in a fully adaptive situation, where the
non-cheating player knew how likely the cheating player was to cheat.
In this case, player~1 was more motivated to cheat than player~2:
The main results are (the values below are the expected payoff to player~1):
\ZbeginE
\item When neither player cheats: $-\zfrac{1}{18}$
\item When player~1 cheats (alone, non-adaptive): $\zfrac{7}{18}$
\item When player~1 cheats (alone, adaptive): $\zfrac{2}{18}$
\item When player~2 cheats (alone, non-adaptive): $-\zfrac{12}{18}$
\item When player~2 cheats (alone, adaptive): $-\zfrac{2}{18}$
\item If both players cheat ``enough'': $0$
\ZendE

The incorporation of cheating detection within Kuhn poker was also
analyzed.
Either players could probabilistically detect if the other player was
cheating.
In this case, when one player always cheats and the other player
detects cheating probabilistically, player~1 gains more by detecting
than player~2 does.
This is due to the fast that Kuhn poker is asymmetric, player~1 is at
a disadvantage by having to bet first.

%--------------------------------
%--------------------------------
\end{document}